\documentclass[superscriptaddress,twocolumn,prl,showpacs]{revtex4}
\usepackage{amsmath,amsfonts,amssymb,bm}

 \newcommand{\cP}{\mathcal{P}} \newcommand{\cF}{\mathcal{F}}
  
\newcommand{\cW}{\mathcal{W}}
    
\begin{document} 
\title{Memory effects in transport through a hopping
insulator: Understanding two-dip experiments} 
\author{V. I. Kozub}
\affiliation{A. F. Ioffe  Physico-Technical Institute of Russian
Academy of Sciences, 194021 St. Petersburg, Russia}
\affiliation{Argonne National Laboratory, 9700 S. Cass Av., Argonne,
IL 60439, USA} 
\author{Y. M. Galperin} 
\affiliation{Department of
Physics, University of Oslo, PO Box 1048 Blindern, 0316 Oslo, Norway}
\affiliation{A. F. Ioffe Physico-Technical Institute of Russian
Academy of Sciences, 194021 St. Petersburg, Russia}
\affiliation{Argonne National Laboratory, 9700 S. Cass Av., Argonne,
IL 60439, USA} 
\author{V. Vinokur} 
\affiliation{Argonne National
Laboratory, 9700 S. Cass Av., Argonne, IL 60439, USA}
\author{A. L. Burin} 
\affiliation{Department of Chemistry, Tulane
University, New Orleans, LA 70118, USA} 
\date{\today}

\begin{abstract} We discuss memory effects in the conductance of
hopping insulators due to slow rearrangements of many-electron
clusters   leading to formation of polarons  close to the electron
hopping sites. An abrupt change in the gate voltage and corresponding
shift of the chemical potential change populations of the hopping
sites, which then slowly relax due to  rearrangements of the
clusters. As a result, the density of hopping  states becomes time
dependent on a scale relevant to rearrangement of  the structural
defects leading to the excess time dependent conductivity.
\end{abstract} 
\maketitle

\paragraph{Introduction  --}

Memory effects in low-temperature transport properties of hopping
insulators have been reported in several systems
\cite{Chorin93,Martinez97,Ovadyahu97,Grenet}.  After excitation from
equilibrium by, e.\, g., a sudden change of a gate voltage, $V_g$, the
conductance of the system increases independent of the sign of the
change. This excess conductance, $\Delta \sigma$, may persist for long
times after the excitation forming the so-called memory cusp,
see~\cite{Ovadyahu06} for a review.

Several concepts were used to explain the  memory cusps in the
dependence of the conductance, $G$, versus the gate voltage,
$V_g$. So-called  \textit{intrinsic} mechanism, is based on the
assumption that the memory effects are due to complex dynamics in the
strongly correlated system of interacting electrons
\cite{Vaknin02,Vaknin98,Muller07}. It is a natural assumption since
hopping insulators lack strong metallic screening and the long-range
Coulomb interaction can be decisive. In~\cite{Muller04} the connection
between the glass-like behavior  and Coulomb gap was argued.  Another
scenario, so-called the \textit{extrinsic}, assumes that the observed
conductance relaxations  are due to the influence of slowly relaxing
atomic configurations acting on the conducting channels. It was first
advocated in~\cite{Adkins84} to explain the occurrence of a $G(V_g)$
cusp in granular gold films.

To the best of our knowledge, the mechanism behind the memory effects
in hopping insulators is far from being fully understood. In
particular,  we are not aware of fully convincing explanation of the
experimental results~\cite{Ovadyahu06} on ``double-dip'' structure of
the $G(V_g)$  dependences, and their relaxations. Recent
experiments~\cite{Ovadyahu07} aimed at studies of the influence of the
sample lateral dimensions on the glassy
properties 
show that there are reproducible conductance
fluctuations having apparently different time scale comparing with the
memory cusps.

 Recently we suggested a simple \textit{extrinsic} model allowing for
the ``two-dip'' behavior of the conductance of a structurally
disordered hopping insulator~\cite{Burin07}. According to this model,
the memory is supported by two-state dynamic structural defects
present in any medium with sufficient amount of structural
disorder~\cite{AHVP}.  The two-level defects get polarized by the
electrons and, in turn, form a \textit{polaron gap} at the hopping
sites decreasing hopping conductance. The slow dynamics of conductance
is then due to slow rearrangement of polaron clouds around the hopping
sites.  An important feature of this extrinsic model qualitatively
explaining logarithmic relaxation and memory
effects~\cite{Ovadyahu97,Grenet}  is presence of a set of fluctuators
possessing dipole moments, uniform density of states and
logarithmically uniform spectrum  of relaxation times. Such
fluctuators can also have intrinsic, electronic nature. Recent
experiments~\cite{Ovadyahu06,Ovadyahu07} can help to determine  which
particular mechanism -- ``extrinsic'' or ``intrinsic'' -- is
responsible for the observed behavior   using the temperature
dependence of the ``typical'' relaxation time defined  by the memory
deeps equilibration. We will show that  the ``chessboard'' electronic
fluctuators  suggested in~\cite{Burin06} to
interpret $1/f$ noise in hopping conductivity also lead   to the
non-equilibrium conductivity behavior similar to that observed
experimentally. As we will see below, this time is sensitive to the
minor deviations  of the relaxation time distribution from its
$\propto 1/\tau$ shape. The model~\cite{Burin06} turns out to be able
to explain the observed \textit{increase} of the typical relaxation
time with temperature, as well as some other observed features.

\paragraph{Main concepts  --}
To calculate the time-dependent conductance of  the system one needs
to know the time-dependent density of hopping states (DOS) at the
Fermi level. The DOS is modified by rearrangement of the populations
of the sites neighboring the sites belonging to the hopping
cluster. This rearrangement driven by hopping electrons lead to
decrease, $U$, of the energy of a hopping site. The polaron shifts $U$
are different for different sites and can be characterized by a
distribution function, $\cF(U,t)$. This distribution slowly depends on
time due to rearrangement of the polaron clouds.

The analysis below is similar to our previous work \cite{Burin07}. A site can be brought from the ground to an excited state by placing
or removing an electron, that can take place only if the excitation
energy, $\varepsilon$, exceeds the polaron gap. Thus the formation of
the polaron gap excludes all the states with $U \ge
\bar{\varepsilon}$. Here $\bar{\varepsilon}$ is the typical excitation
its equilibrium value being the width of the hopping band,
$\varepsilon_h$. For an equilibrium state created by sudden change
$\delta V_g \gg \varepsilon_h$ of the gate voltage $\bar{\varepsilon}
\approx \delta V_g$. Thus, $\bar{\varepsilon} \sim
\max\{\varepsilon_h, \delta V_g\}$. The  effect of the polaron cloud
on the conductivity can be  estimated as
\begin{equation}\label{eq:ans2} 
\frac{\delta G (t)}{G} \sim -
\int^{\infty}_{\bar{\varepsilon}} {\cal F}(U,t)\, d U\, .
\end{equation}  
This result is straightforward -  the sites inside the
polaron cloud cannot be occupied, and the density of states is
\textit{less} than the universal one. Thus the relative decrease in
the density of states due to polarons can be estimated as a relative
volume occupied by the polaron cloud.

To find $\cF(U,t)$ one need to specify the slow relaxing aggregates
producing the polaron shift $U$.  Following  previous studies of
1/f noise in hopping conductivity~\cite{Burin06}  we assume that slow
dynamics is due to chess-board electro-neutral clusters having $2N$
sites placed at a distance $\approx r$ between each other.
The relaxation rate, $\tau$ of a given aggregate  depends both on the
number of sites, $2N$, in the aggregate and on the typical distance,
$r$, between the sites, and energy difference $E$ between its lowest
energy levels.  We are interested in the domain where $E \lesssim T$
since in the opposite case the cluster resides in its ground
state. Let us define the distribution $P(N,r,E)$ such that $P(N,r,E)\,
dN\,dr\, dE$ is the number of the clusters per unit volume having the
parameters within the region $(N+dN, r+dr, E+dE)$. To estimate this
distribution, we will take into account that the typical aggregate
volume is $Nr^3$ while the energy bandwidth for small energies is
equal to $\sqrt{N} e^2/\kappa r$. Thus, the \textit{total} density of
states, ${\cal W}(N,r,E)$,  of all aggregates with the intersite
distances $r'$ larger than some $r$ is given as  $\cW (N,\rho,E) \sim
\lambda^N/T_0N^{3/2}a^3 \rho^2$. 
%
Here  $\lambda$ is a probability to
add additional pair of sites  to the aggregate; $\rho \equiv r/a$
where $a$ is the localization length, while $T_0\equiv e^2/\kappa
a$. In this way we get the expression for the partial density as $P(N,
\rho) \sim \lambda^N/T_{0} N^{3/2}a^3\rho^2$.
%
%
Since we are interested in the case $E \lesssim T \ll T_0$ the partial
density is $E$-independent.  Let us first consider the
Efros-Shklovskii (ES) regime~\cite{Shklovskii-Efros} of  variable
range hopping (VRH). In this case the relaxation rate for cluster
rearrangement can be expressed by the interpolation
formula~\cite{Burin06}
\begin{equation}
  \label{eq:001a} 
\tau^{-1}(N,\rho)
=\nu_0\left(e^{-N^{2/3}\xi^2/\rho}+e^{-N\rho}\right), 
\end{equation} 
$\xi \equiv (T_0/T)^{1/2}$.
The first contribution corresponds to formation of a
``domain wall" in the aggregate, while the second one corresponds to
coherent tunneling transitions leading to re-charging of all aggregate
sites. The distribution of relaxation rates can be calculated as
\begin{equation}
  \label{eq:100} \cP(\tau)=\int dN\, d\rho\,
P(N,\rho)\delta\left[\tau(N,\rho)-\tau\right] \, .
\end{equation} As shown in \cite{Burin06}, the integral is dominated
by the values $N=N_c$, $\rho=\rho_c$, where
\begin{equation}
  \label{eq:101} N_c(\tau)=[\ln (\nu_0\tau)/\xi]^{6/5}\, , \
\rho_c(\tau) = \xi/[N_c(\tau)]^{1/6}\, .
\end{equation} The quantities $N_c$ and $\rho_c$ characterize most
important clusters among those switching during the time $\sim
\nu^{-1}$. Since the number of electrons in cluster $N_{c}$ depends 
logarithmically on all relevant parameters and it cannot be very large because other relevant parameters are exponentially sensitive to it we set $N_{c}\sim 1$ following Ref. \cite{Burin06}.
Substituting
(\ref{eq:101}) in Eq.~(\ref{eq:100}) 
we obtain 
\begin{equation}
  \label{eq:t-dist} \cP(\tau)\sim \frac{P_0}{\tau}\,
\frac{1}{(\nu_0\tau)^\alpha 
}\, .
\end{equation} Here $P_0 \equiv 1/(T_0a^3\xi^3)$, $\alpha (\tau) \sim
\xi^{-6/5}\ln^{1/5}(\nu_0\tau) \ll 1$.

Note that at $\delta V_g \gg \varepsilon_h$ the typical distance $R
\approx e^2/(\kappa \delta V_g)$ corresponding to the polaron shift
$\sim \delta V_g$ produced by a nearest neighbor  turns out to be less
than the hopping length $r_h = a\xi$, as well as a typical size $\sim
r_h 
$ of the aggregate. Consequently one can treat the
interaction between the site belonging to the percolation cluster and
a fluctuator as a contact one. Thus, $U(R) \sim e^2/ \kappa R$ where
$R$ is the distance between the hopping site and its nearest neighbor
belonging to the fluctuator. As  result, the contribution of the
clusters with relaxation time $\tau$ to the distribution of polaron
shifts is
\begin{equation}
\cF_\tau (U) =\frac{8\pi R^2
\cP
(
\tau)}{d\ln U/dR}= \frac{8 \pi e^6 }{\kappa^3
U^3}
\cP(
\tau).
\end{equation}
 Here we have taken into account that only
aggregates with $E < U$ form the polaron as well as the fact that each
of the 2N cites of the aggregate can be coupled to the hopping
site. The proper distribution $\cF(U,t)$ is determined by the
manipulation protocol. For example, if the system is brought to some
state at time $t_0$ by a sudden change of the gate voltage then the
polaron clouds are formed by all the fluctuators which have changed
their states by the observation time, $t$. Consequently,
$\cF(U,t)=\int_{t_0}^t \cF_\tau (U)\, d\tau$.

\paragraph {Discussion --}
By now we were discussing the ES regime of VRH. One can expect that
the number of the metastable aggregates  strongly decreases within the
Mott regime. Indeed, aggregates are constructed from the sites where
the intersite Coulomb energies are of the order of single-particle
energies. This is not the case for typical hopping sites in the Mott
VRH regime where the spread in the energies of the localized states is
rather due to extrinsic disorder than to Coulomb interaction. Because
of this spread it is less probable to find a set of sites forming a
two-state aggregate. The addition factor entering the probability
$\lambda$ for adding a pair of sites to an aggregate can be estimated
as the ratio of the width of the Coulomb gap, $\Delta_C \sim
T_0^{3/2}T_M^{-1/2}$, to the typical hopping band in the Mott regime,
$\varepsilon_M =T_M^{1/4}{T}^{3/4}$. Here $T_M$ is the characteristic
temperature of the Mott VRH, $\sigma \propto e^{-(T_M/T)^{1/4}}$; we
have defined $\Delta_C$ as the temperature of crossover between
regimes of ES and Mott VRH. The ratio $\Delta_C/\varepsilon_M
=(\Delta_C/T)^{3/4}$ is additional factor entering the probability
$\lambda$. Deeply in the Mott regime, $T \gg \Delta_C$, this factor is
small. Since $T_M \propto 1/g_0$ where $g_0$ is the Mott density of
states $\Delta$ decreases with decrease of $g_0$.

To compare our prediction with experimental results
of~\cite{Ovadyahu97,Ovadyahu06,Vaknin02,Vaknin98,Ovadyahu07,Zvi07,Vaknin00}
obtained using InO films we assume that in the absence of Coulomb
interaction their DOS, $g_0$, would be energy independent at the
energies less or of the order of both room temperature,
$T_{\text{r}}$, and the shift in the chemical potential, $\delta \mu$,
due to variation in the gate voltage, $\delta V_g$, within the
dip. This DOS consists of localized   and extended states split by the
mobility edge, $\varepsilon_{\text{m}}$. The closer is $\mu$ to
$\varepsilon_{\text{m}}$, the larger is the localization length and,
consequently, the hopping conductance.  Different samples have
different $g_0$ and different $\mu$ with respect to
$\varepsilon_{\text{m}}$. This picture is  conventional for materials
with large amount of disorder.

The authors of~\cite{Vaknin98} determine the carrier concentration
from the Hall coefficient at $T=T_{\text{r}}$. Then the found
concentration, $n_{\text{r}}$,  is just the  concentration of the
extended carriers, and one can estimate DOS as $g_0 \approx
n_{\text{r}}/T_{\text{r}}$. The width of the dip is related with the
shift in chemical potential as $\delta V_g = g_0\, \delta \mu/C
\approx \delta \mu$,
where $C$ is capacitance. Thus the width of the dip is $\propto g_0$
and, for a given position of the chemical potential, it is correlated
with $n_{\text{r}}$ found in~\cite{Vaknin98}. On the other hand, the
samples with the same  $g_0$, but different positions of the chemical
potential, have different resistance -- the lower $\mu$ the lower is
localization length and the larger is  resistance. From
phenomenological point of view, the samples with the same $g_0$, but
larger resistance can be characterized as ``more dirty''.
For density of states $g_0 = (4 \cdot 10^{18}\, \mathrm{cm}^{-3})/(300
\, \mathrm{K}) \sim 10^{32}$ cm$^{-3}$erg$^{-1}$ which roughly
corresponds to the threshold of the memory effect in
\cite{Vaknin98,Ovadyahu06}, the Coulomb gap can be estimated as
$\varepsilon_C = (g_0 e^6 /\kappa^3)^{1/2} \sim 3 \cdot 10^{-15}$ erg.
At the same time, $T=4$ K corresponds to a crossover between the
Mott and  ES VRH regimes, and at this temperature $\xi \sim
5$ and  $\varepsilon_h \sim \varepsilon_C$.  Thus, we explain the
observed in~\cite{Vaknin98,Ovadyahu06} correlation between pronounced
decrease of the memory effect and decrease of the dip width  by decrease of the density of the metastable aggregates in the
Mott regime.

According to Eq.~(\ref{eq:ans2}) the relative change of the
conductance is $\propto \bar{\varepsilon}^{-2}$. Thus the shape of the
dip can be cast in an interpolation formula
\begin{equation}
  \label{eq:int-form} \frac{\delta G}{G} \sim -\frac{(e^2/\kappa)^3
P_0\,Q(t)}{(C \delta V_g/g_0)^2+\varepsilon_h^2}\, , \ Q(t)
\equiv \int\frac{d\tau\, (\nu_0 \tau)^{-\alpha}}{\tau  
}
\end{equation} 
where the limits of integration are determined by the
manipulation protocol. The temperature dependence of the dip magnitude
is given by the product $\xi^{-9/5}\varepsilon_h^{-2} \propto
T^{-0.1}$.  Showing the same trend  as in experiment~\cite{Ovadyahu06}
it is still much weaker.  Assuming $\ln \nu_0\tau \sim 20$ we estimate
the height of the dip as  $\xi^{1/5} (\ln \nu_0 \tau)^{-6/5} \sim
0.05$ that  is in agreement with experiments.

To analyze time dependence of the dip we take into account that the
parameters of the system (like concentration, localization length,
etc.) are somewhat different for different gate voltage. It is clearly
demonstrated by the fact that the $G(V)$ curve have a systematic slop
(subtracted in course of studies of dip).  Let us for simplicity
assume that the bonding parameter $\xi$ depends on the gate voltage,
say, through the localization length.  As it seen from experiment, $G$
increases with increase of concentration, i.\,e. with
$V_g$. Consequently, we can assume that   $\xi$ decreases with
increase of $V_g$. Correspondingly, the parameters of the aggregates
also depend on $V_g$ and are different for the aggregates responsible
for different dips in the two dip experiment.

In the well known double-dip experiments~\cite{Ovadyahu06} the
typical relaxation time is defined as following. First,
the gate voltage in a gated sample is rapidly changed from some
initial to some final value $V_{g1}$ (we have denoted the time of this
variation as $\tau_{\min}$). Then it  is kept
constant until some time, which we will denote as $t_{\max}$. During
this time the conductivity slowly (apparently logarithmically)
decreases to some value, $G_0 -\delta G (t_{\max},
\tau_{\min})$. Then the gate voltage is swept to some other value,
$V_{g2}$ and kept constant, the conductivity decreasing with time
forming a new dip, $G_0 -\delta G (t, \tau_{\min})$.
Here with a logarithmic accuracy we ascribed the same estimate
$\tau_{\min}$ for the fast process of switching from $V_{g1}$ to
$V_{g2}$. Let us assume that the shift of $\mu$ due
to the variation of $V_g$ is less than the typical
single-particle energies of the sites forming the aggregates. Then
at the new value of $V_g$ the aggregates responsible for the polaron
gap at $V_g = V_{g1}$ stay at the same configuration of the
occupation numbers. However the occupation of the sites forming the
percolation cluster at $V_g = V_{g1}$ at $V_g = V_{g2}$ is
completely changed provided that the shift of the chemical potential
is larger than $\varepsilon_h$. Thus the aggregates responsible for
the first dip start to relax. However at the times $t < \tau_{\max}$
the slow aggregates still preserve the configuration corresponding
to $V_g = V_{g1}$; thus the first dip is partly restored if the gate
voltage is returned to the value $V_{g1}$. The depth of the restored
dip at the time $t$ is expected to be $\delta G (\tau_{\max},
t)$. The relaxation time, $\bar{\tau}$, is defined according to
equality of the depths of the ``old" and ``new" dips. This condition
corresponds to the equality
$\bar{\tau}$ is
calculated according to the following procedure:
\begin{equation}
\int_{\tau_{\min}}^{\bar{\tau}} 
\cP(\tau)\, d\tau =
\int_{\bar{\tau}}^{\tau_{\max}}
\cP(\tau)\,  d\tau \, .
\end{equation}
 Since 
$\cP(\tau)\approx \text{const}\times \tau^{-1}(\nu_0\tau)^{-\alpha}$ Eq. (\ref{eq:t-dist}) then up to the lowest
approximation in $\alpha$,
$\bar{\tau}_0=\sqrt{\tau_{\max}\tau_{\min}}$. Next iteration provides
the correction $\bar{\tau}_1/\bar{\tau}_0
=-\tfrac{\alpha}{8}\ln^2(\tau_{\max}/\tau_{\min})$, which leads to
\textit{decrease} of $\bar{\tau}$ with temperature increase. However,
it is
more sensitive to the possible dependence of the
parameter $\alpha$ on the \textit{gate voltage}. The first (initial)
dip and the second one correspond  to different gate voltages one can
expect that they are formed by the states with different localization
lengths. Consequently, the values of $\alpha$ are different. Denoting
them as $\alpha_{1,2}$ for the first and the second dips,
respectively, and assuming that  $|\alpha_1-\alpha_2| \ll \alpha$  we
arrive at the second temperature-dependent correction
$\bar{\tau}_2/\bar{\tau}_0 \approx [(\alpha_1-\alpha_2)/4\alpha]
\ln(\nu_0\tau_0)(\tau_{\max}/\tau_{\min})^{\alpha/2}$.  As follows
from experiments~\cite{Vaknin98,Vaknin00}  the conductance for the
second dip is larger than for the first one, which indicates the
smaller value of $\xi$ for the second dip. Since $\alpha \propto
\xi^{-6/5}$ one concludes that $\alpha_1>\alpha_2$.  Thus the
correction $\bar{\tau}_2$ is \textit{positive} and increases with
temperature both due to increase of $\alpha$ and $\nu_0$.  This trend
can qualitatively explain observed weak increase of $\bar{\tau}$ with
temperature.  Similar conclusion can be made for another
protocol~\cite{Ovadyahu06,Zvi07} for determining a typical relaxation
time.
As we have seen, increase of resistance, or $\xi$ is correlated with
decrease of the exponent $\alpha$ and subsequent slowing down the time
evolution. This can be a qualitative explanation of the observed in~\cite{Ovadyahu97} slowing down the time evolution with increase of
disorder.

Now let us discuss an effect of external magnetic
field~\cite{Ovadyahu06}.  One can imagine  two possible mechanisms:
(i) shrinkage of the wave function manifesting itself as a positive
addition $\propto H^{2}$ to the hopping
exponent~\cite{Shklovskii-Efros}; (ii) spin effect related to a
presence of doubly occupied centers. In the latter case the magnetic
field align spins of singly occupied sites which blocks the
spin-conserving hops between the singly occupied sites. One can expect
that  at $\mu g H << \varepsilon_h$ the latter mechanism leads to a
positive magnetoresistance, its magnitude being proportional to
relative contribution of the doubly occupied sites.  Both mechanisms
can be accounted for by  a field-dependent  increase of the tunneling
length $\rho$ entering the second item in the r.h.s. of
Eq.~(\ref{eq:001a}). That would,in turn, lead to decrease of 
the exponent $\alpha$.

In the experiments \cite{Ovadyahu97}, the observed magnetoresistance
is only weakly dependent on the magnitude of resistance and decreases
with resistance increase. This behavior seems to be contradictory to
the wave shrinkage mechanism since in that case the magnetoresistance
would dramatically increase with increase of the hopping exponent
$\xi$.  In addition, the shrinkage effect is expected to be small for
materials with small localization length.  Thus, it is the spin
mechanism that probably dominates.

Interestingly, the double-dip memory effects are not observed in
standard semiconductor materials. We believe that the reason is that
the ES regime of VRN in such materials either occurs at very low temperatures
(less than few Kelvins) which implies weak heat
withdrawal, or corresponds to very large  resistances. Both of these
factors seems to be 
disadvantageous for typical memory experiments.

To conclude, our model qualitatively explains the following
experimentally observed features  of the memory effect: (i) double-dip
behavior of the conductance as a function of gate voltage; (ii)
suppression of the above phenomenon at small carrier concentrations
due to possible crossover to the Mott regime of VRH;  (iii) rather
counter-intuitive slowing down of the time evolution (expressed
through the effective relaxation time $\bar{\tau}$) with temperature
increase; (iv) slow power-law relaxation tending to logarithmic with
increase of resistance; (v) qualitative dependences of memory dips on
temperature and electron concentration; (vi) slowing down the
relaxation with increasing of external magnetic field and degree of
disorder characterized by increase of resistance.

\acknowledgments
The work was supported by the U.~S.~Department of Energy Office of Science
through contract No. DE-AC02-06CH11357, by Norwegian Research
Council through the USA-Norway Bilateral Program,   
and Tulane University Research and Enhancement Program.


\end{document}